\documentclass[twocolumn]{article}
\usepackage[pdftex]{graphicx}
\usepackage{multirow} 
\usepackage[normalem]{ulem}
\usepackage{float}
\usepackage{units}
\usepackage[caption=false,font=footnotesize]{subfig}
\usepackage{stfloats}
\usepackage[margin=0.9in]{geometry}

\usepackage[pdftex]{graphicx,xcolor}
\usepackage[fleqn]{amsmath}
\usepackage{newtxtext}
\usepackage[varg]{newtxmath}
\usepackage{algorithm,algpseudocode}
\usepackage{multirow}
\usepackage{ulem}
\usepackage{graphicx}       
\DeclareGraphicsExtensions{.pdf,.png,.jpg,.jpeg}
\usepackage{cite}
\usepackage{xspace}
\newcommand{\itutp}{\textit{P.1203}\xspace}

\newcommand{\tran}{\textit{Tran's}\xspace}
\newcommand{\guo}{\textit{Guo's}\xspace}
\newcommand{\vriend}{\textit{Vriendt's}\xspace}
\newcommand{\yin}{\textit{Yin's}\xspace}
\newcommand{\rehman}{\textit{Rehman's}\xspace}
\newcommand{\cqm}{\textit{CQM}\xspace}
\newcommand{\cqmtran}{\textit{CQM+Tran's}\xspace}
\newcommand{\cqmguo}{\textit{CQM+Guo's}\xspace}
\newcommand{\cqmvri}{\textit{CQM+Vriendt's}\xspace}
\newcommand{\cqmyin}{\textit{CQM+Yin's}\xspace}
\newcommand{\cqmitutp}{\textit{CQM+P.1203}\xspace}

\newcommand{\CQMQoMEX}{\textit{preCQM}\xspace}
\newcommand{\addFist}[1]{\textcolor{black}{#1}}

\usepackage[pdftex]{graphicx}
\usepackage{ragged2e}
\usepackage{array}
\newcolumntype{R}[1]{>{\raggedright \arraybackslash}m{#1}}
\newcolumntype{C}[1]{>{\centering \arraybackslash}m{#1}}
\newcolumntype{?}{!{\vrule width 1pt}}
\usepackage[usestackEOL]{stackengine}

\usepackage{xcolor,cite,etoolbox}
\makeatletter
\pretocmd\@bibitem{\color{black}\csname keycolor#1\endcsname}{}{\fail}
\newcommand\citecolor[1]{\@namedef{keycolor#1}{\color{blue}}}
\makeatother
\citecolor{}

\usepackage{authblk}
\usepackage[colorlinks=true, urlcolor=black]{hyperref}
\begin{document}
	
	\title{Cumulative Quality Modeling for HTTP Adaptive Streaming}
	\author[1]{Huyen~T.~T.~Tran}
	\author[2]{Nam~Pham~Ngoc}
	\author[3]{Tobias~Ho{\ss}feld}
	\author[3]{Michael Seufert}
	\author[1]{Truong~Cong~Thang}
	\affil[1]{The University of Aizu, Aizuwakamatsu, Japan}
	\affil[2]{Vin-University project, Vietnam}
	\affil[3]{University of W{\"u}rzburg, Germany}
	\date{}                     %% if you don't need date to appear
	\setcounter{Maxaffil}{0}
	\renewcommand\Affilfont{\itshape\small}
	
	\maketitle
\begin{abstract}
HTTP Adaptive Streaming has become the de facto choice for multimedia delivery nowadays. However, the quality of adaptive video streaming may fluctuate strongly during a session due to throughput fluctuations. So, it is important to evaluate the quality of a streaming session over time. In this paper, we propose a model to estimate the cumulative quality for HTTP Adaptive Streaming. In the model, a sliding window of video segments is employed as the basic building block. Through statistical analysis using a subjective dataset, we identify three important components of the cumulative quality model, namely the minimum window quality, the last window quality, and the average window quality. Experiment results show that the proposed model achieves high prediction performance and outperforms related quality models. In addition, another advantage of the proposed model is its simplicity and effectiveness for deployment in real-time estimation. Our subjective dataset as well as the source code of the proposed model have been made available to the public. 
\end{abstract}

\section{Introduction}\label{Section:Introduction}

HTTP Adaptive Streaming (HAS) has become the de facto choice for multimedia delivery nowadays. In HAS, a video is encoded into different quality versions. Each version is further divided into a series of segments. Depending on throughput fluctuations, segments of appropriate quality versions will be delivered from the server to the client, which results in quality variations during a session. Therefore, a key challenge in HAS is how to evaluate the quality of a session over time. The evaluation can provide service providers with suggestions to enhance the quality of services. Also, some existing studies deploy quality models to build and evaluate effective adaptive streaming strategies~\cite{HAS_TCSVT}. 

Here, we would like to differentiate three concepts of the quality as follows. 
\begin{itemize}
	\item \textit{Continuous quality} means the instantaneous quality which is continuously perceived at any moment of the session.
	\item \textit{Overall quality} means the quality of a whole session.  
	\item \textit{Cumulative quality} means the quality cumulated from the beginning up to any moment of the session. Obviously, the concept of overall quality is a special case of cumulative quality.
\end{itemize}
It should be noted that the concepts of continuous quality and overall quality have been mentioned in Recommendation ITU-R BT.500-13 and ITU-T P.880~\cite{ITU2004-P880,ITU2012-BT500} and have been investigated in a large number of previous studies.

To the best of our knowledge, however, few previous studies have actually considered the cumulative quality.
%In~\cite{QoE_tobias2011_memory}, the cumulative quality was investigated in the context of Web services. 
The work in~\cite{QoE_ZWangTimevarying} was the first study on the cumulative quality of a video streaming session, where the authors focused on the impact of quality variations. However, this work employed very short sessions, only \unit[5--15]{seconds}. 

In this study, our goal is modeling the cumulative quality of HTTP adaptive video streaming. We first carry out a subjective test to measure the cumulative quality of long sessions of 6 minutes. Then, the impacts of quality variations, primacy, and recency are investigated. Based on the obtained results, a cumulative quality model (called \cqm) is proposed. In the proposed model, a sliding window of video segments is the basic unit of computation. It should be noted that, in the following, the term "window" means either the conceptual sliding window or a window at a certain location. Experiment results show that the quality of the last window, the average window quality, the minimum window quality, and the maximum window quality are key components of the cumulative quality model. Also, it is found that the proposed model outperforms seven existing models. Moreover, the proposed model is applicable to real-time quality monitoring thanks to its low computation complexity. To the best of our knowledge, the proposed model is the first cumulative quality model for actual streaming sessions. 

The remainder of this paper is organized as follows. Section~\ref{Sec:Related} discusses the related work and our contributions. Because the proposed model is based on an analysis of subjective results, the subjective test is presented in Section~\ref{Sec:Subjective}. Then, Section~\ref{Sec:Model} presents the proposed cumulative quality model. In Section~\ref{Sec:ModelEval}, we evaluate the performance and computation complexity of the proposed model and compare it to seven existing models. Finally, conclusions are drawn in Section~\ref{Sec:Conclu}.

\section{Related work and contributions}\label{Sec:Related} 
In this section, we will discuss the work related to three types of quality, namely, 1)~continuous quality, 2)~overall quality,  and 3)~cumulative quality. Also, our contributions in this study will be presented at the end of this section.

\subsection{Continuous quality}\label{Subsec:RelatedConti}

The recommendation ITU-R BT.500-13 describes the Single Stimulus Continuous Quality Evaluation (SSCQE) method for subjective assessment of the continuous quality. In this method, test sessions are displayed in a random order. Each subject, while watching a video, is asked to continuously move a slider along a continuous scale so that its position reflects his/her selection of quality at that instant. All subjects' quality ratings at each instant of each video are averaged to compute a mean opinion score (MOS) of that instant.     

The work in~\cite{QoE_Chen2014} is the first study on the continuous quality of a streaming session. Note that, in this paper, the authors use the term "time-varying quality" to refer to "continuous quality". To measure the continuous quality, the authors conducted a subjective test similar to the SSCQE method. Then, a continuous quality model is proposed, taking into account the impact of the recency. In particular, a Hammerstein-Wiener model was employed to predict the continuous quality of 5-minute long sessions. As this work is focused on continuous quality, the model mainly depends on the quality values of the last 15 seconds. %Based on experiment results, it was found that the continuous quality at an instant mainly depended on the segment quality values during last 15 seconds //13 segments. 

\cite{QMon_shafiq_infocom18} uses machine learning to predict initial delay, stalling, and video quality from the network traffic in windows of \unit[10]{s}. The considered features are derived from IP or TCP/UDP headers only. ViCrypt~\cite{QMon_seufert2019_stream} detects QoE degradations on encrypted video streaming traffic in real-time within \unit[1]{s} by using a stream-like analysis approach with two continuous sliding windows and a cumulative window. The features are based on packet-level statistics of the network traffic, and allow to accurately recognize initial delay and stalling~\cite{QMon_seufert2019_stream}, as well as video resolution and the average bitrate~\cite{QMon_wassermann2019_let}.

\cite{texas2018} presents a continuous quality predictor using an ensemble of Hammerstein-Wiener models, while~\cite{QoE_BampisRecurrent2018ML,QoE_EswaraLSTM} developed  neural-network-based continuous quality models. As discussed in Recommendation ITU-R BT.500-13~\cite{ITU2012-BT500}, the continuous quality values of a session can be utilized to obtain the overall quality. However, this issue is currently under study~\cite{ITU2012-BT500,QoE_bovikTimeVarying2011,QoE_bampis2017Dataset}. 

%In~\cite{QoE_BampisRecurrent2018ML}, a neural network based continuous quality model was proposed for short streaming sessions of 72 seconds. Similar to~\cite{QoE_Chen2014}, the authors in~\cite{QoE_BampisRecurrent2018ML} conducted a subjective test using the SSCQE method to measure the continuous quality. The model also considers the recency effect. In particular, the inputs of the model consist of the quality values of 15 previous segments, 15 previous playback statuses, and the time distance since the last video impairment~\cite{QoE_bampis2017continuous}. 

%As discussed in Recommendation ITU-R BT.500-13~\cite{ITU2012-BT500}, the continuous quality values of a session can be utilized to obtain the overall quality. However, this issue is currently under study~\cite{ITU2012-BT500,QoE_bovikTimeVarying2011,QoE_bampis2017Dataset}. 

\subsection{Overall quality}\label{Subsec:RelatedOverall}
The overall quality perceived by the end-users can be quantified with the concept of Quality of Experience (QoE). In terms of video streaming, the QoE states to what extent users are annoyed or delighted with the provided streaming~\cite{qualinet2013qoe,QoE_texas2019}. In ~\cite{QoE_tobias2013_DTMA}, it was found that the impact of the initial delay of the video stream is not severe, whereas the impact of stalling, i.e., playback interruptions, is significant. To model the impact of the interruptions, previous studies generally used some statistics such as the number of interruptions~\cite{QoE_singh2012qualityML, QoE_liu2015_deriving}, the average~\cite{QoE_singh2012qualityML}, the maximum~\cite{QoE_singh2012qualityML}, the sum~\cite{QoE_rodriguez2016_video, QoE_liu2015_deriving}, and the histogram~\cite{tran2016_GC} of interruption durations. To ensure a smooth streaming when end-users face throughput fluctuations, e.g., in mobile networks, HAS allows to adapt the video bit rate to the network conditions. Thereby, initial delay and stalling can be reduced, which are severe QoE degradations of video streaming. However, due to the bit rate adaptation, the visual quality of the video might vary, which introduces an additional QoE factor, called quality variations~\cite{QoE_tavakoli2016_JSAC}.

Existing studies on overall quality were mostly limited to short sessions (about \unit[1--3]{minutes})~\cite{QoE_bellLab2013_QoEmodel,QoE_ywang2015_assessing,tran2016_GC,QoE_tobias2014_assessing}. These studies mainly focused on the impact of the quality variations. This impact is generally modeled by some statistics of segment quality values and switching amplitudes (i.e., differences between consecutive segment quality values) such as average~\cite{ QoE_bellLab2013_QoEmodel}, standard deviation~\cite{QoE_bellLab2013_QoEmodel}, minimum~\cite{QoE_ywang2015_assessing}, median~\cite{QoE_ywang2015_assessing}, histogram~\cite{tran2016_GC}, and time duration on different quality levels~\cite{QoE_tobias2014_assessing}. 

For long sessions, the primacy and recency are also important factors to be considered. Here, the primacy (recency) factor refers to the influences of quality degradations near the beginning (end) of a session. The authors in~\cite{QoE_tavakoli2016_JSAC} found that the primacy and recency both have significant impacts on the overall quality of a session.~\cite{QoE_Seufert2013_pool} studies different temporal pooling methods, which emphasize different aspects (e.g., recency, lowest quality), for aggregating objective quality metrics into an overall quality score. In~\cite{QoE_rodriguez2016_video}, the authors proposed an overall quality model, taking into account the impacts of the quality variations, primacy, and recency. Specifically, a session is divided into three temporal intervals. In each interval, the impact of quality variations is modeled by the frequencies of switching types. Each switching type is defined based on resolutions and frame rates. To take into account the impact of the primacy and recency, each interval is simply assigned a weight to represent its contribution to the overall quality of the session. The experiment results then revealed that the first interval has the highest weight, and so the largest contribution to the overall quality.

In the latest stage of ITU-T P.1203~standardization for quality assessment of streaming media, a model (called \itutp) is recommended for predicting the overall quality, where session durations are from 1 to 5 minutes~\cite{ITU1203_3}. The \itutp model also takes into account the impacts of quality variations, primacy, and recency. %Similar to~\cite{QoE_rodriguez2016_video}, a session is divided into three temporal intervals regardless of the session's duration. %CHECK: https://github.com/itu-p1203/itu-p1203/blob/master/itu_p1203/rfmodel.py (only considered in RF part, which is only 25% of overall score. For this RF part, the split in three parts for video scores occurs in function scale_moses, but only the first third is treated differently (else clause) from the other part (if clause) --> I would omit this discussion
Then, to model the impact of quality variations, the authors used the average of the segment quality values in each temporal interval and various statistics calculated over a whole session, such as the total number of quality direction changes and the difference between the maximum and minimum segment quality. To take into account the impact of the primacy and recency, the authors used a weighted sum of all segment quality values in the session. 

%Besides the above mentioned factors, some existing studies investigated the impacts of other factors such as initial delay and interruptions to the overall quality~\cite{QoE_tobias2013_DTMA,QoE_tobias2012_initial,QoE_singh2012qualityML,QoE_liu2015_deriving,QoE_rodriguez2016_video,QoE_seufert2015_survey}. In ~\cite{QoE_tobias2013_DTMA,QoE_tobias2012_initial}, it was found that the impact of the initial delay is not severe, whereas the impact of the interruptions is significant. To model the impact of the interruptions, previous studies generally used some statistics such as the number of interruptions~\cite{QoE_singh2012qualityML,QoE_liu2015_deriving}, the average~\cite{QoE_singh2012qualityML}, the maximum~\cite{QoE_singh2012qualityML}, the sum~\cite{QoE_rodriguez2016_video,QoE_liu2015_deriving}, and the histogram~\cite{tran2016_GC} of interruption durations. 

\subsection{Cumulative quality} \label{Subsec:RelatedCumm}

To the best of our knowledge, the only previous study on the cumulative quality of a streaming session is in~\cite{QoE_ZWangTimevarying}, where the authors presented some qualitative observations regarding the impact of quality variations. However, the authors employed simple simulated sessions of very short durations (\unit[5--15]{seconds}) with only \unit[1--3]{segments}. It is found that, when there is a quality variation with a small switching amplitude, the cumulative quality is quite stable. Meanwhile, a large switching amplitude results in a significant change of the cumulative quality. From these observations, the authors proposed a cumulative quality model, in which a piecewise linear function of switching amplitudes was used to quantify the impact of the quality variations.  

The preliminary work of our cumulative quality research was presented in~\cite{tran2018_QoMEX}. In this paper, the previous work is extended significantly in several aspects. First, we carried out more subjective tests with new videos and so the dataset is now doubled. Second, factors in the model are extensively studied with one-way analysis of variance (ANOVA). Third, different window sizes are analyzed and used for different window quality statistics. Fourth, the model performance is explored in detail and the best setting is recommended. Finally, the evaluation is extended with more related models and in-depth analysis of models' performances with respect to the length of sequences as well as models' computation complexity.  

The contributions of our work have two general categories. First, we build a dataset that is specific to the cumulative quality. Our dataset helps to investigate how existing overall quality models perform cumulative quality prediction. Second, we propose a new cumulative quality model that can well predict the cumulative quality of streaming sessions. In particular, the distinguished features of our study are as follows. 
\begin{itemize}
	\item 
	First, a subjective test was specifically designed for measuring the cumulative quality of HAS sessions. In our test, there are in total 72 test sequences generated from six 6-minute long videos. The total time required for rating these sequences was approximately 160 hours.
	\item 
	Second, through statistical analysis, insights into the impacts of three factors of quality variations, primacy, and recency are provided. In particular, it is found that the impacts of the quality variations and recency are significant. However, no significant impact of the primacy is observed. 
	\item 
	Third, we proposed a new cumulative quality model that takes into account the impacts of the quality variations and recency. Experiment results show that the proposed model is able to predict well the cumulative quality of streaming sessions.
	\item 
	Fourth, a comparison of the proposed model with seven existing models was conducted. This is the first time a large number of quality models have been investigated for cumulative quality prediction. Experiment results show that the proposed model outperforms the existing models. 
	\item 
	Fifth, it was found that the proposed model is applicable to real-time quality monitoring thanks to its low computation complexity. This feature is especially important for cost-effective evaluation of streaming technologies.
	
\end{itemize}  

\section{Subjective Test for Cumulative Quality }\label{Sec:Subjective}

In this study, to measure the cumulative quality over time, each streaming session was converted into test sequences of different lengths. In the test, each subject viewed a random sequence and then rated the quality of the whole sequence. This approach is similar to that used in~\cite{QoE_ZWangTimevarying}, where each 15-second long session was divided into three sequences of 5, 10, and 15~(seconds).

\begin{table}[t]
	\caption{Features of Source Videos}
	\label{tab:DescripSVideos}
	\centering
	\resizebox{\columnwidth}{!}{
	\begin{tabular}{|p{0.45in}|p{2.5in}|p{1.5in}|} \hline
		\centering\textbf{Video} 
		& \centering\textbf{Content} 
		& \multicolumn{1}{c|}{\textbf{Type}} \\ 
		\hline 
		Video \#1 & Slow movements of characters & Animated video, Movie \\ 
		\hline 
		Video \#2 & A story about Sintel and her friend, a dragon. & Animated video, Movie \\ 
		\hline 
		Video \#3 & Conversations of characters & Natural video, Movie \\ 
		\hline
		Video \#4 & A talk show host analyzing news & Natural video, News \\ 
		\hline
		Video \#5 & A documentary about the science experiment & Natural video, Documentary \\ 
		\hline
		Video \#6 & A soccer match & Natural video, Sport \\ 
		\hline
	\end{tabular}}
\end{table}%

\begin{table}[t]
	\caption{Average Bitrates of Versions}
	\label{tab:BR}
	\centering
	\resizebox{\columnwidth}{!}{%
		\begin{tabular}{|c|c|c|c|c|c|c|}
			\hline
			\multirow{2}{*}{\textbf{Version}} & \multicolumn{6}{c|}{\textbf{Average bitrate (kbps)}} \\ \cline{2-7}
			& \textit{\textbf{Video \#1}} & \textit{\textbf{Video \#2}} & \textit{\textbf{Video \#3}}&\textit{\textbf{Video \#4}} & \textit{\textbf{Video \#5}} & \textit{\textbf{Video \#6}}\\ \hline
			\textbf{1} & 146  & 187  & 187   & 179 & 455 & 570        \\ \hline
			\textbf{2} & 196  & 239  & 244   & 310 & 794 & 1034       \\ \hline
			\textbf{3} & 310  & 333  & 353   & 382 & 1010 & 1304      \\ \hline
			\textbf{4} & 455  & 482  & 528   & 548 & 1397 & 1823      \\ \hline
			\textbf{5} & 717  & 717  & 813   & 675 & 1764 & 2295      \\ \hline
			\textbf{6} & 1118 & 1097 & 1263  & 791 & 2017 & 2647      \\ \hline
			\textbf{7} & 1751 & 1743 & 2005  & 977 & 2549 & 3330      \\ \hline
			\textbf{8} & 2802 & 2910 & 3362  & 1303 & 3209 & 4382     \\ \hline
			\textbf{9} & 4538 & 4993 & 6089  & 1613 & 3930 & 5500     \\ \hline
		\end{tabular}%
	}
\end{table}

There are in total six 6-minute long videos used in this study, denoted by Video \#1, Video \#2, Video \#3, Video \#4, Video \#5, and Video \#6, with features presented in Table~\ref{tab:DescripSVideos}. These videos were encoded using H.264/AVC (libx264) with a frame rate of \unit[24]{fps}. In this study, we used two adaptation sets, each consisted of 9 versions with different QP values and/or resolutions. In particular, the 9 versions in the first adaptation set have the same resolution of 1280$\times$720 and 9 different QP values of 52, 48, 44, 40, 36, 32, 28, 24, and 20. The first adaptation set was used to generate the streaming sessions of Video \#1, Video \#2, and Video \#3. The 9 versions in the second adaptation set are different in both resolution and QP. Specifically, the 9 versions correspond to 9 combinations of QP values and resolutions of \{24, 256$\times$144\}, \{26, 426$\times$240\}, \{24, 426$\times$240\}, \{26, 640$\times$360\}, \{24, 640$\times$360\}, \{26, 854$\times$480\}, \{24, 854$\times$480\}, \{26, 1280$\times$720\}, \{24, 1280$\times$720\}. The second adaptation set was used to generate the streaming sessions of Video \#4, Video \#5, and Video \#6. The average bitrates of the versions are shown in Table~\ref{tab:BR}. In this study, every version is divided into short segments with the duration of \unit[1]{second}.  

\begin{figure}[t]
	\centering
		\hfil
		\includegraphics[width=0.8\linewidth, keepaspectratio=true]{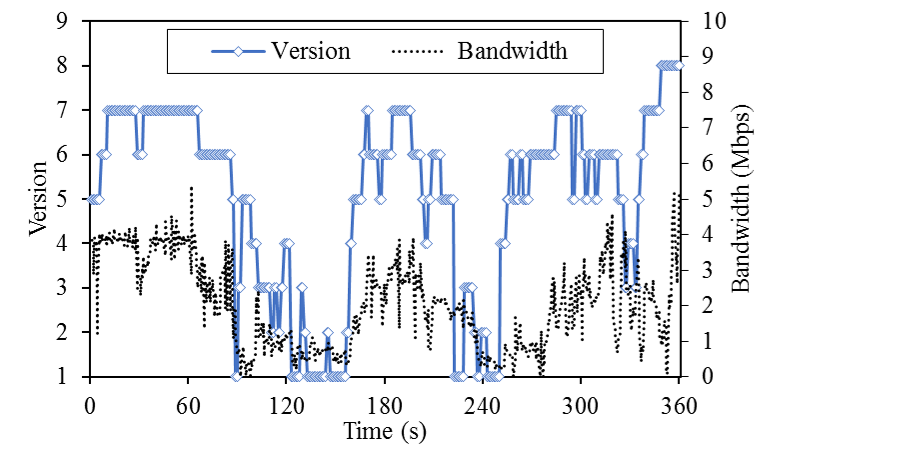}
		\caption{An example of version variations in a streaming session.}
		\label{fig:ExVar}
\end{figure}

For each video, two full-length sessions of \unit[6]{minutes} were generated by using the adaptation method of~\cite{thang2013_JCN} and  two bandwidth traces from a mobile network~\cite{HAS_muller2012_evaluation}. The duration of \unit[6]{minutes} was selected such that it is longer than the average video duration watched on YouTube, which is \unit[5:01]{minutes}~\cite{QoE_nam2016_qoewhycat}. The bandwidth traces have average throughputs varying from \unit[1484.87]{kbps} to \unit[3432.33]{kbps}, and standard deviations from \unit[867.01]{kbps} to \unit[1252.75]{kbps}. An example of version variations in a 6-minute session is provided in Fig.~\ref{fig:ExVar}. 

From each full-length session, six test sequences were extracted, from the time-stamp 0 to the 1$^{st}$, 2$^{nd}$, 3$^{rd}$, 4$^{th}$, 5$^{th}$, and 6$^{th}$~minute. So, from the six original videos, there were in total 72 test sequences, with durations from 1 minute to 6~minutes. The total duration of all the test sequences is \unit[252]{minutes}. Because a rating time which is longer than \unit[1.5]{hours} may cause fatigue and boredom~\cite{P.9132014}, the subjective test was divided into four parts that were conducted in different days. The duration of each part was approximately \unit[1.5]{hours}, of which about 1~hour was spent for rating the test sequences. In the rating process, every \unit[20]{minutes}, there was a break of \unit[10]{minutes}. In order to avoid boredom, each subject took part in at most two test parts. 

The subjective test was conducted using the absolute category rating (ACR) method. Test conditions were designed following Recommendation ITU-T~P.913~\cite{P.9132014}. In the subject-training stage, the subjects got used to the procedure and the range of quality impairments. In the test, the sequences were randomly displayed on a black background. The screen has the size of 14 inches and a resolution of 1366$\times$768. Given a sequence, each subject gave a score at the end of the sequence with the value ranging from 1 (worst) to 5 (best), which reflects his/her option of quality of the whole sequence. 

There were in total 71 subjects taking part in the test. The total time of the test was approximately \unit[160]{hours}. Screening analysis of the test results was performed following Recommendation ITU-T~P.913~\cite{P.9132014}, and two subjects were rejected. After discarding these subjects' scores, each test sequence was rated by 23 valid subjects. The MOS of each sequence was computed as the average of the valid subjects' scores. 

\begin{figure}[t]
	\begin{center}
		\includegraphics*[width=0.8\linewidth, keepaspectratio=true]{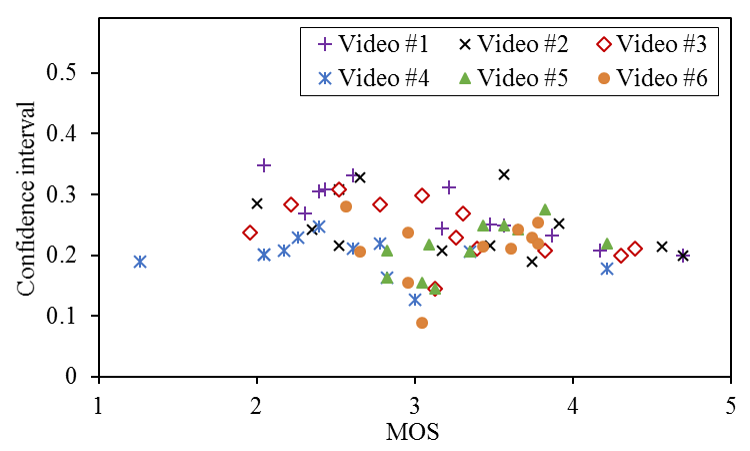}
		\caption{95\% confidence intervals of MOSs.}
		\label{fig:CI}
	\end{center}
\end{figure}

The 95\% confidence intervals of the MOSs are shown in Fig.~\ref{fig:CI}. In general, the confidence intervals are in the range 0.08 to 0.35. Also, the MOSs are in the range from 2 to about 4.7. This means the cumulative quality varies drastically during a session.  %The average of the confidence interval widths is 0.22 over all test sequences.  

\section{Cumulative Quality Model}\label{Sec:Model}

\begin{figure}[t]
	\begin{center}
		\includegraphics*[width=0.8\linewidth, keepaspectratio=true]{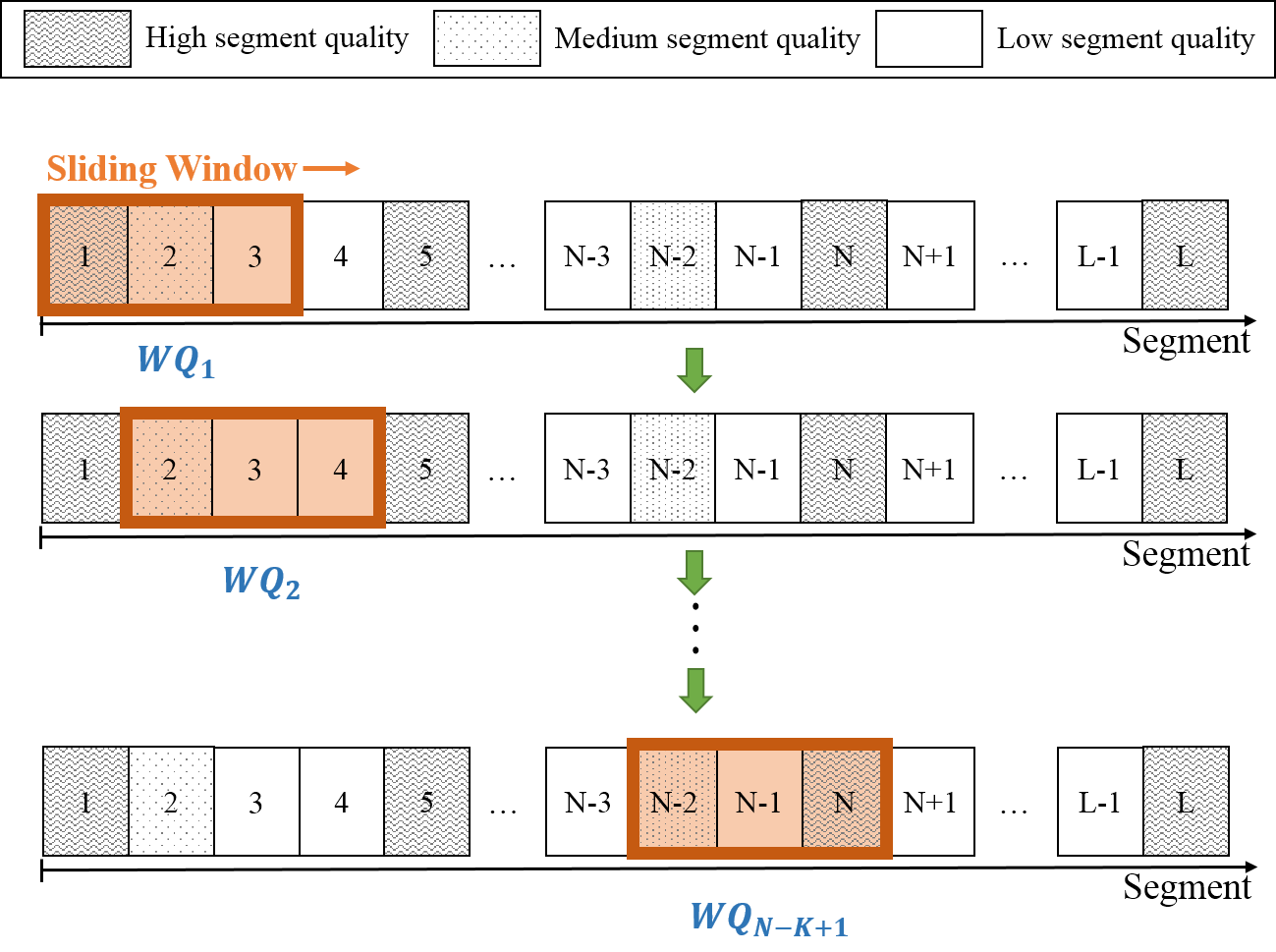}
		\caption{An illustration of \textit{"sliding window"} with size $K$=3}
		\label{fig:WinCon}
	\end{center}
\end{figure}

\subsection{Overview}
To build a cumulative quality model taking into account the impacts of multiple factors, the basic ideas of our solution are as follows. 

\begin{itemize}
	\item Quality variations over a long session are divided into long-term and short-term changes. Specifically, short-term changes refer to quality variations of neighboring segments, while long-term changes refer to quality variations between temporal intervals. 
	
	\item To represent the impact of long-term changes, the concept of "sliding window" is used. Specifically, a window of $K$ segments is moved along the session, segment by segment  as illustrated in Fig.~\ref{fig:WinCon}. After each time, a window quality value is computed. 
	
	\item To represent the impact of short-term changes within a window, an existing overall quality model is used. For this purpose, such a model is called \textit{window quality model}.  
	
	\item The cumulative quality value at any time point is computed based on window quality values, taking into account the impacts of factors such as long-term changes and recency. Note that, at the first time points, when the watched video duration is (very) short (i.e., less than $K$ segments), the corresponding cumulative quality values are directly computed from the window quality model.  
\end{itemize}   

In the next subsection, effect analysis of the quality variations, primacy, and recency will first be presented. Then, based on the obtained results, a cumulative quality model will be proposed. 	

\subsection{Proposed quality model}
As mentioned, to identify the key components of a cumulative quality model, we carried out a statistic analysis of some window quality values. In particular, the first window quality value $WQ^K_f$ and the last window quality value $WQ^K_l$ were employed to represent the impacts of the primacy and recency respectively. For the factor of long-term changes, four window quality statistics are considered, which are the average window quality $WQ^K_{av}$,  the maximum window quality $WQ^K_{ma}$, and the minimum window quality $WQ^K_{mi}$ of all windows until a given time point.

Suppose that the window is just moved to the $N^{th}$ segment with $N \ge K$. By using the window quality model, the window quality value $WQ^K_{N-K+1}$ is calculated. After that, the window quality statistics of $WQ^K_f$, $WQ^K_l$, $WQ^K_{av}$, $WQ^K_{ma}$, and $WQ^K_{mi}$ are updated by the following equations.
\begin{equation}
WQ^K_f = WQ^K_1.
\end{equation}
\begin{equation}
WQ^K_l = WQ^K_{N-K+1}.
\end{equation}
\begin{equation}
WQ^K_{av} = \begin{matrix}
WQ^K_1, &  \text{if}\  N = K \\ 
\frac{WQ^K_{av} \times (N-K) + WQ^K_{N-K+1}}{N-K+1}, &  \text{otherwise}.
\end{matrix} 
\end{equation}
\begin{equation}
WQ^K_{mi} = \begin{cases}
WQ^K_1, &  \text{if}\  N = K \\ 
\min \{WQ^K_{mi},WQ^K_{N-K+1}\}, &  \text{otherwise}.
\end{cases}
\end{equation}	
\begin{equation}
WQ^K_{ma} = \begin{cases}
WQ^K_1, &  \text{if}\  N = K \\ 
\max \{WQ^K_{ma},WQ^K_{N-K+1}\}, &  \text{otherwise}.
\end{cases}
\end{equation}
%\begin{align}
%WQ^K_{std} = \begin{cases}
%WQ^K_1, &  \text{if}\  N = K \\ 
%\sqrt{\frac{1}{N-K}\sum_{i=1}^{N-K+1}\left | {WQ^K}_i-{WQ^K}_{av} \right |}, &  \text{otherwise}.
%\end{cases}
%\end{align}

Table~\ref{tab:AnovaParam} shows the obtained results from one-way analysis of variance (ANOVA). To assess the effect size, partial Eta-squared values ($\eta_p^2$) are also reported in Table~\ref{tab:AnovaParam}. Here, the window size $K$ is set from 10 to 60~seconds with the step size of 10~seconds. The window quality model is \addFist{our previous} overall-quality model~\cite{tran2016_GC} (called \tran), \addFist{which is found to be very effective in representing the impact of short-term changes~\cite{tran2016_GC}.} 

\begin{table*}[t]
	\caption{Results of Effect Analysis of Window Quality Statistics}
	\label{tab:AnovaParam}
	\begin{center}
		%\resizebox{\columnwidth}{!}{%
		\begin{tabular}{|c|c|c|c|c|c|c|c|}
			\hline
			\multicolumn{2}{|c|}{\multirow{2}{*}{\textbf{\begin{tabular}[c]{@{}c@{}}Window quality\\ statistics\end{tabular}}}} & \multicolumn{6}{c|}{\textbf{\begin{tabular}[c]{@{}c@{}}Window size $K$ (seconds)\end{tabular}}}                                          \\ \cline{3-8} 
			\multicolumn{2}{|c|}{}                                                                                              & \textit{\textbf{10}} & \textit{\textbf{20}} & \textit{\textbf{30}} & \textit{\textbf{40}} & \textit{\textbf{50}} & \textit{\textbf{60}} \\ \hline
			\multirow{3}{*}{$\textit{\textbf{WQ}}^K_f$}                                    & \textit{\textbf{F}}                                & 4.868                & 1.594                & 8.589                & 2.088                & 7.321                & 1.478                \\ \cline{2-8} 
			& \textit{\textbf{p}}                                & 0.027                & 0.207                & 0.003                & 0.149                & 0.007                & 0.224                \\ \cline{2-8} 
			& \textit{\textbf{$\eta_p^2$}}                                & 0.003                & 0.001                & 0.005                & 0.001                & 0.004                & 0.001                \\ \hline
			\multirow{3}{*}{$\textit{\textbf{WQ}}^K_l$}                                    & \textit{\textbf{F}}                                & 2.111                & 0.149                & 6.959                & 6.687                & 18.977               & 16.063               \\ \cline{2-8} 
			& \textit{\textbf{p}}                                & 0.146                & 0.699                & 0.008                & 0.010                & <0.001               & <0.001               \\ \cline{2-8} 
			& \textit{\textbf{$\eta_p^2$}}                                & 0.001                & 0.000                & 0.004                & 0.004                & 0.010                & 0.008                \\ \hline
			\multirow{3}{*}{$\textit{\textbf{WQ}}^K_{av}$}                                 & \textit{\textbf{F}}                                & 4.103                & 9.359                & 0.207                & 1.404                & 11.613               & 44.283               \\ \cline{2-8} 
			& \textit{\textbf{p}}                                & 0.043                & 0.002                & 0.649                & 0.236                & <0.001               & <0.001               \\ \cline{2-8} 
			& \textit{\textbf{$\eta_p^2$}}                                & 0.002                & 0.005                & 0.000                & 0.001                & 0.006                & 0.023                \\ \hline
			\multirow{3}{*}{$\textit{\textbf{WQ}}^K_{mi}$}                                  & \textit{\textbf{F}}                                & 3.826                & 2.202                & 3.338                & 16.730               & 38.648               & 6.397                \\ \cline{2-8} 
			& \textit{\textbf{p}}                                & 0.051                & 0.138                & 0.068                & <0.001               & <0.001               & 0.012                \\ \cline{2-8} 
			& \textit{\textbf{$\eta_p^2$}}                                & 0.002                & 0.001                & 0.002                & 0.009                & 0.020                & 0.003                \\ \hline
			\multirow{3}{*}{$\textit{\textbf{WQ}}^K_{ma}$}                                  & \textit{\textbf{F}}                                & 12.075               & 0.896                & 1.971                & 16.366               & 19.644               & 6.958                \\ \cline{2-8} 
			& \textit{\textbf{p}}                                & <0.001               & 0.344                & 0.161                & <0.001               & <0.001               & 0.008                \\ \cline{2-8} 
			& \textit{\textbf{$\eta_p^2$}}                                & 0.006                & 0.000                & 0.001                & 0.009                & 0.010                & 0.004                \\ \hline
			
		\end{tabular}
		%}
	\end{center}
\end{table*}%

The $p$ values in Table~\ref{tab:AnovaParam} indicate that, for all the considered window sizes, no significant effect was observed for $WQ^K_f$ (i.e., $p>0.001$). In contrast, significant results with small effects were obtained for $WQ^K_l$ (i.e., $p<0.001$ and $\eta_p^2<0.06$) when the window size $K$ is 50 or 60~seconds. Especially, the larger effect size was found for the window size of 50~seconds (i.e., $\eta_p^2=0.010$ vs. $\eta_p^2=0.008$). This implies that the impact of the primacy on the cumulative quality can be neglected, while the impact of the recency has to be considered. 

With regard to long-term changes, some significant effects with small sizes were also observed for $WQ^K_{av}$, $WQ^K_{mi}$, and  $WQ^K_{ma}$ (i.e., $p<0.001$ and $\eta_p^2<0.06$). Particularly, the window size corresponding to the strongest effect size is 60~seconds for $WQ^K_{av}$ (i.e., $\eta_p^2=0.023$), 50~seconds for $WQ^K_{mi}$ (i.e., $\eta_p^2=0.020$), and 50~seconds for $WQ^K_{ma}$ (i.e., $\eta_p^2=0.010$). This implies that the three window quality statistics of the average, minimum, and maximum quality should be considered. 

To sum up, the results suggest that $WQ^K_l$, $WQ^K_{av}$, $WQ^K_{mi}$, and $WQ^K_{ma}$ should be key components of a cumulative quality model. Based on these observations, we propose a cumulative quality model which is given by	
\begin{equation}\label{eqn:05}
%\text{\resizebox{0.5\textwidth}{!}{$
CQM=w_1\cdot WQ^{50}_l  +w_2\cdot WQ^{60}_{av} +w_3\cdot WQ^{50}_{mi} +w_4\cdot WQ^{50}_{ma}, 
%$}}
\end{equation}
where $w_1, w_2, {w_3},$ and ${w_4}$ are the corresponding weights of $WQ^{50}_l$, $WQ^{60}_{av}$, $WQ^{50}_{mi}$, and $WQ^{50}_{ma}$ components.

It is interesting to note that the proposed model is in agreement with the peak-end rule~\cite{QoE_Kahneman1993}. The peak-end rule says that users judge an experience largely at its peak and at its end. Here the peaks (lowest and highest) of a session are the minimum window quality $WQ^{50}_{mi}$ and the maximum window quality $WQ^{50}_{ma}$. The end is the significance of the recency effect $WQ^{50}_l$ shown in our model. In the case of speech quality, a large impact of the minimum quality on QoE was also shown~\cite{QoE_Koster2017}. For HTTP adaptive streaming, it was found in~\cite{QoE_tobias2014_assessing} that the number of quality switches is not statistically significant, but the time the video is played out on each quality level. Also~\cite{QoE_Seufert2013_pool} showed that a good temporal pooling method is taking the average over the whole session, implying that $WQ^{60}_{av}$ is a key influence factor. Thus all the key factors of the proposed model are inline with the findings in previous studies. Yet, the \cqm model is the first one that integrates these factors into a single model for predicting the cumulative quality of HAS sessions.

In the next section, we will investigate the performance of the proposed model and some existing models.

\section{Model Evaluation and Analysis}\label{Sec:ModelEval}

\subsection{Evaluation Methodology}\label{Subsec:EvalMetho}

This section is divided in two evaluations, each aiming at an important question. In the first evaluation, we will investigate what is the best window quality model for the proposed model. The second evaluation is carried out to see if existing overall quality models can predict cumulative quality, especially in long sessions. 

There are in total seven existing models employed in this study, which are denoted by \tran~\cite{tran2016_GC}, \guo~\cite{QoE_ywang2015_assessing}, \vriend~\cite{QoE_bellLab2013_QoEmodel}, \yin~\cite{QoE_Yin_2015}, \itutp~\cite{ITU1203_3,ITUT_implement1,ITUT_implement2,ITUT_implement3},  \rehman~\cite{QoE_ZWangTimevarying}, and \CQMQoMEX~\cite{tran2018_QoMEX}. Note that the \CQMQoMEX model is one proposed in our preliminary work~\cite{tran2018_QoMEX}. Among these models, only the \rehman and \CQMQoMEX models were proposed for cumulative quality prediction, the other models were originally proposed for overall quality prediction. 

Similar to~\cite{QoE_Database_ZDuanmu2018, QoE_QoEIndex_ZDuanmu2018}, to evaluate the performance of existing models, we implemented the models using the parameter settings stated in the corresponding publications. In addition, following Recommendation ITU-T P.1401~\cite{ITUT_Rec1401}, a first order linear regression between predicted scores and MOSs was performed for each model to compensate for possible variances between subjective tests. The obtained coefficients of slope and intercept will be stated in the following subsections.

For the evaluations, the 72 sequences in our dataset were randomly divided into two sets, namely a training set of 36 sequences and a test set of 36 remaining sequences. The division is repeated 50 times, resulting in 50 pairs of training and test sets. The training set was used to obtain the model parameters by curve fitting. The test set was to evaluate the performance of the models.  

In order to measure the performance of the models, we used two metrics of Pearson Correlation Coefficient (PCC) and Root-Mean-Squared Error (RMSE). The PCC and RMSE values reported below were averaged over the 50 test sets. Since the capability of real-time processing is an especially important feature for cumulative quality models, we also measured computation complexity of the models. In this study, the computation complexity was measured as the average time required to obtain a cumulative quality value per 1-second long segment. The measurement was conducted on a computer with Intel Core i3-2120 processor at 3.30GHz and 8GB RAM.   

\subsection{Performance Analysis of \cqm Model}\label{Subsec:OptimalSet}

In this subsection, we first investigate the performance of the proposed model using different window quality models. Our goal is to find the best window quality model for the proposed model. Then, the model parameters are determined based on result analysis.

%\begin{figure}[t]
%	\begin{center}
%		\hfil
%		\includegraphics*[width=0.6\linewidth, keepaspectratio=true]{WS.png}
%		\caption{Performance of \cqm model using different window sizes}
%		\label{fig:WS}
%	\end{center}
%\end{figure}

%\begin{table}[t]
%	\caption{Model Parameters of Proposed Model}
%	\label{tab:Param}
%	\begin{center}
%		\begin{tabular}{|C{0.6in}|C{0.6in}|C{0.6in}|}\hline 
%			\multicolumn{3}{|c|}{\textbf{Parameter}} \\ \hline 
%			${w_1}$ & ${w_2}$ & ${w_3}$ \\ \hline 
%			0.29 & 0.31 & 0.40 \\\hline 
%		\end{tabular}
%	\end{center}
%\end{table}%

\subsubsection{Window quality model}\label{Sec:CompareWQM}
In this part, the five overall quality models of \tran, \guo, \vriend, \yin, and \itutp are employed to obtain window quality values. Note that these models all take into account the impact of short-term changes. Further, note that \rehman and \CQMQoMEX are cumulative models, which were not used here, but is only used later for comparison purpose.

\begin{table}[t]
	\caption[]{Performance of \cqm Model using Different Window Quality Models}
	\label{tab:ShortModels}
	\begin{center}
		\begin{tabular}{|R{0.83in}|C{0.3in}|C{0.3in}|C{0.3in}|C{0.3in}|}\hline 
			\centering \multirow{3}{*}{\textbf{Model}} & \multicolumn{4}{c|}{\textbf{Performance}} \\ \cline{2-5}
			{}&\multicolumn{2}{c|}{\textbf{~Training set}} & \multicolumn{2}{c|}{\textbf{Test set}} \\ \cline{2-5}
			{} & PCC & {\shortstack[c]{RMSE}} & PCC & {\shortstack[c]{RMSE}} \\ \hline 
			\cqmtran   		&0.93&	0.26&	0.93&	0.27 	\\ \hline
			\cqmguo   		&0.91&	0.30&	0.90&	0.31 	\\ \hline
			\cqmvri			&0.92&	0.27&	0.92&	0.28 	\\ \hline
			\cqmyin   		&0.91&	0.29&	0.91&	0.29 	\\ \hline
			\cqmitutp   	&0.91&	0.30&	0.91&	0.29	\\ \hline
		\end{tabular}
	\end{center}
\end{table}%

Table~\ref{tab:ShortModels} shows the performance of the \cqm model using the different window quality models. It can be seen that the performance of the \cqm model is generally good with all the window quality models. Especially, the combination \cqmtran provides the best prediction performance. Specifically, the values of PCC and RMSE are 0.93 and 0.26 for the training set, and 0.93 and 0.27 for the test set. The main reason is that, for modeling the impact of short-term changes, \tran model utilizes the histograms of segment quality values and switching amplitudes which are shown to be more effective  than the statistics used in the remaining models~\cite{tran2017_IEICEhistogram}. 

\begin{figure*}[b]
	\hrulefill
	\begin{align}\label{EqCQM} 
	CQM&=w_1\cdot WQ^{50}_l  +w_2\cdot WQ^{60}_{av} +w_3\cdot WQ^{50}_{mi} +w_4\cdot WQ^{50}_{ma}\\
	&=0.280\cdot WQ^{50}_l  +0.426\cdot WQ^{60}_{av} +0.280\cdot WQ^{50}_{mi} +0.014\cdot\label{Par} WQ^{50}_{ma}.  			
	\end{align}
\end{figure*}

Since the combination \cqmtran provides the best performance, \tran model is used as the window quality model in the rest of this paper.

%\subsubsection{Window size}
%In this part, the performance of the \cqm model is evaluated using different window sizes. As mentioned, \tran model is employed to obtain window quality values. Fig.~\ref{fig:WS} shows the performance of the proposed model with different window sizes ranging from 2~to 90~seconds with the step size of 2~seconds. 
%It is clear that, given a window size, the training set always achieves higher PCC values and lower RMSE values than that of the test set. In general, the behaviors of the PCC and RMSE curves for the training and test sets are similar. In particular, the prediction performance improves quickly (i.e., PCC value increases quickly while the RMSE value drops quickly) when the window size is increased to 14 seconds. When the window size is from 14 to 50 seconds, some small improvements are observed for the PCC and RMSE. The best prediction performance for the test set is achieved with the window size of 50 seconds. Specifically, the PCC and RMSE values are 0.94 and 0.26 for the training set, and 0.93 and 0.27 for the test set. When the window size increases beyond 50~seconds, the PCC falls sharply and the RMSE rises dramatically. Therefore, to achieve the highest performance, the window size should be 50~seconds. In the rest of this paper, this value of the window size will be used.

\subsubsection{Model parameters} Similar to~\cite{QoE_liu2015_deriving}, in order to obtain the model parameters, we pick the (best) training set, which provides the \addFist{best performance}. The \addFist{cumulative quality model} is given by (\ref{EqCQM}).

%Table~\ref{tab:Param} shows the parameters $w_1, w_2,$ and ${w_3}$, which correspond to the best performance.
The \addFist{positive} numerical values of the weights $w_1, w_2, w_3,$ and ${w_4}$ reconfirm the observations in Section~\ref{Sec:Model} that $WQ^K_l$, $WQ^K_{av}, WQ^K_{mi}$, and $WQ^K_{ma}$ are key components of the cumulative quality model. Also, the impacts of the quality variations and recency are significant on the cumulative quality of a session. In addition, it can be seen that ${w_2}$ is highest while ${w_4}$ is lowest. So the impact of the average window quality is strongest, and the impact of the maximum window quality is weakest.

\subsection{Model Comparison}\label{Subsec:Compare}
In this subsection, we compare the \cqm model and the seven existing models in terms of the prediction performance and the computation complexity. 

\begin{figure}[t]
	\begin{center}
		\includegraphics[width=0.8\linewidth, keepaspectratio=true]{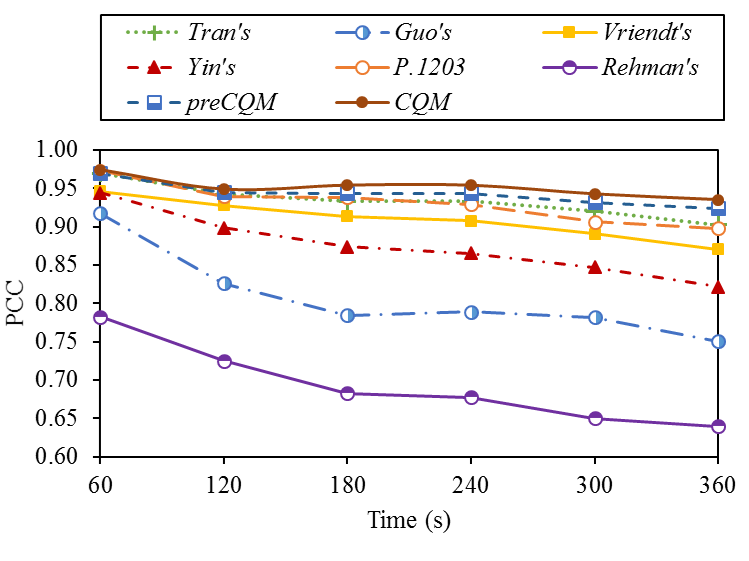}
		\caption{Performance of models with different sequence lengths}
		\label{fig:PerUpLen}
	\end{center}
\end{figure}

Fig.~\ref{fig:PerUpLen} shows the PCC values of the models  with different sequence lengths. We can see that, when the sequence length is 1 minute, the PCC values of \tran, \guo, \vriend, \yin, and \itutp models are high (i.e., PCC $\ge 0.92$). This suggests that these models can predict well the overall quality of a short session, and thus each of them can be used as a window quality model with good performance as discussed in Subsection~\ref{Sec:CompareWQM}. 

However, when the sequence length increases, the PCC values of the models decrease. Among the models, the PCC of the \cqm model is highest for all the sequence lengths. Meanwhile, the performance of \rehman model is lowest. A possible explanation is that \rehman model is designed using very short sessions with a duration of \unit[5--15]{seconds}. Thus it is not really suitable for longer sessions (i.e., \unit[1--6]{minutes}). In addition, there is no consideration for long-term changes and recency in \rehman, \tran, \guo, \vriend, and \yin models, so the performances of these models are all lower than that of the \cqm model. \addFist{It turns out that the simple \CQMQoMEX model's performance is only a little worse than that of the \cqm model. It is mainly because the impact of maximum window quality is actually small in the obtained \cqm model.}

%Although both the \itutp and \cqm models take into account the factors of quality variations and recency, the difference in their performances becomes larger when the sequence length increases. In the \itutp model, to represent long-term changes, a session is heuristically divided into three intervals, and the interval quality values are then used as inputs for a machine learning module. With respect to the recency, its impact is modeled by a weighted sum of segment quality values. The fixed division of three intervals could be the reason for its lower performance. 

\begin{table}[t]
	\caption{Performance of Models in Predicting Cumulative Quality }
	\label{tab:RefModels}
	\begin{center}
		\resizebox{\columnwidth}{!}{
		\begin{tabular}{|l|c|c|c|c|c|}\hline 
			\multirow{3}{*}{\centering \textbf{~~~Model}} &  \multicolumn{2}{c|}{\multirow{2}{*}{\textbf{Coefficients}}} & \multicolumn{2}{c|}{\multirow{2}{*}{\begin{tabular}{c}\textbf{Performance}\\\textbf{(Test set)}\end{tabular}}} &\multirow{2}{*}{\centering {\shortstack[c]{\textbf{Computation}\\\textbf{complexity} \\ \textbf{(ms)} }} } \\ 
			&\multicolumn{2}{c|}{} &  \multicolumn{2}{c|}{}  &\\ \cline{2-5}
			&  \textbf{\textit{Slope}} & \textbf{\textit{Intercept}} &\textbf{\textit{PCC}} & \textbf{\textit{RMSE}} &\\ \hline
			\tran				&1.24				&-1.27			&0.90	&0.31		&0.22 \\ \hline
			\guo				&1.01				&-0.25			&0.74	&0.48		&0.02 \\ \hline		
			\vriend				&1.02				&-0.41			&0.86	&0.36		&0.04 \\ \hline
			\yin				&1.07				&-0.79			&0.81	&0.41		&0.06 \\ \hline
			\itutp				&1.04				&-0.93			&0.89	&0.32		&1722.78 \\ \hline
			\rehman				&25.11				&-26.68			&0.63	&0.55		&0.05 \\ \hline
			\CQMQoMEX        	&\textemdash	    &\textemdash	&0.92	&0.32		&0.17 \\ \hline
			\textbf{\cqm}    	&\textemdash	    &\textemdash	&\textbf{0.93}	&\textbf{0.27}		&\textbf{0.17} \\ \hline %0.62
		\end{tabular}}
	\end{center}
\end{table}%

\begin{figure}[t]
	\begin{center}
		\centering
		\includegraphics[width=0.8\linewidth, keepaspectratio=true]{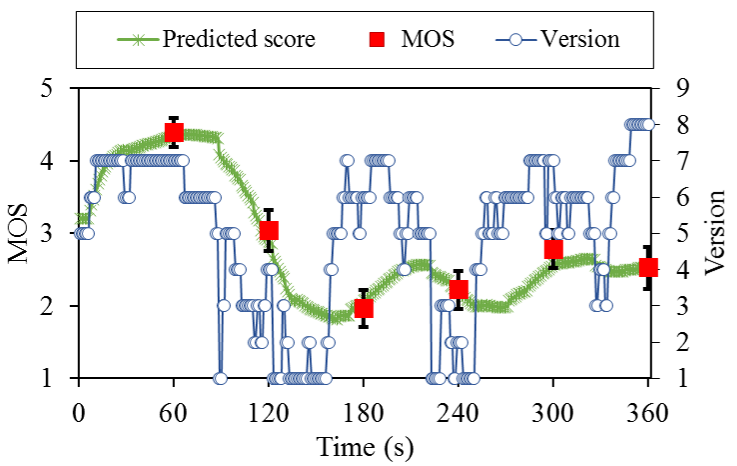}
		\caption{An example of the cumulative quality values of a streaming session.}
		\label{fig:Dis}
	\end{center}
\end{figure}

Table~\ref{tab:RefModels} summarizes the performances and the computation complexity of the models. Here, the PCC and RMSE are averaged over the 50 test sets containing sequences of different lengths. We can see that the results of performances are similar to those in Fig.~\ref{fig:PerUpLen}. In particular, the performance of the \cqm model is highest and the performance of \rehman model is lowest.  

Regarding the computation complexity, it can be seen that the \cqm model takes less than 1ms to obtain a cumulative quality value, and so the cumulative quality can be updated after every segment as the window slides forward. In other words, the \cqm model is applicable to real-time quality monitoring. \addFist{It should be noted that the reported computation complexity of the \cqm model  is in case where different window sizes are computed in parallel.}

For the \itutp model, its computation complexity is considerably higher than the others. In particular, the \itutp model takes an average of 1.72s to calculate a cumulative quality value. Meanwhile, the remaining models have an average processing time less than 1ms per cumulative quality value. 

To better understand the cumulative quality, Fig.~\ref{fig:Dis} shows the MOSs and the predicted scores by the \cqm model corresponding to the adaptation result in Fig.~\ref{fig:ExVar}. We can see that the predicted scores closely follow to the MOSs. In addition, the cumulative quality fluctuates strongly during the session. This means that evaluating the overall quality at the end of a streaming session is obviously not enough to fully understand the quality of the video streaming service. So, cumulative quality over time is of crucial importance in adaptive streaming.

\section{Conclusions and Future Work}\label{Sec:Conclu}

In this paper, we have presented a model for predicting the cumulative quality of adaptive video streaming. The proposed model was developed based on the concept of a "sliding window" over a streaming session, where each window is characterized by a quality value. 

First, a subjective test was specifically designed and conducted for measuring the cumulative quality. Second, through statistical analysis, it was found that the impacts of the quality variations and recency are significant. We integrated the significant key components, namely,  the last window quality, the average window quality, the minimum window quality, and the maximum window quality, into a new cumulative quality model \cqm, which is able to accurately predict the cumulative quality of streaming sessions. The advantage of the proposed \cqm model is its simplicity, while being inline with other well known effects from literature, namely, the applicability of simple temporal pooling plus the peak-end rule.

The \cqm model was compared with seven existing models, where it could outperform the other models in predicting the cumulative quality. Moreover, the proposed model is applicable to real-time quality monitoring thanks to its low computation complexity. This feature is especially important for cost-effective evaluation of streaming technologies, e.g., for real-time quality monitoring of video streams. In the future, the model will be used to assess the quality of different adaptive streaming techniques. Also, we will develop novel quality adaptation strategies, which are based on the \cqm model.

%It was shown that the average window quality, the minimum window quality, and the last window quality are the key components of the cumulative quality model. The advantage of the proposed \cqm model is its simplicity, while being inline with other well known effects from literature, namely, the applicability of simple temporal pooling plus the peak-end rule.
%Moreover, the model can be well applied for real-time quality monitoring. In future work, we will employ the model to evaluate typical adaptation strategies in the literature. 	

% Generated by IEEEtran.bst, version: 1.14 (2015/08/26)
%\bibliographystyle{IEEEtran}
%\bibliography{IEEEabrv,Citation_Stand_full}

\begin{thebibliography}{10}
	\providecommand{\url}[1]{#1}
	\csname url@samestyle\endcsname
	\providecommand{\newblock}{\relax}
	\providecommand{\bibinfo}[2]{#2}
	\providecommand{\BIBentrySTDinterwordspacing}{\spaceskip=0pt\relax}
	\providecommand{\BIBentryALTinterwordstretchfactor}{4}
	\providecommand{\BIBentryALTinterwordspacing}{\spaceskip=\fontdimen2\font plus
		\BIBentryALTinterwordstretchfactor\fontdimen3\font minus
		\fontdimen4\font\relax}
	\providecommand{\BIBforeignlanguage}[2]{{%
			\expandafter\ifx\csname l@#1\endcsname\relax
			\typeout{** WARNING: IEEEtran.bst: No hyphenation pattern has been}%
			\typeout{** loaded for the language `#1'. Using the pattern for}%
			\typeout{** the default language instead.}%
			\else
			\language=\csname l@#1\endcsname
			\fi
			#2}}
	\providecommand{\BIBdecl}{\relax}
	\BIBdecl
	
	\bibitem{HAS_TCSVT}
	S.~{Cical\`{o}}, N.~{Changuel}, V.~{Tralli}, B.~{Sayadi}, F.~{Faucheux}, and
	S.~{Kerboeuf}, ``Improving qoe and fairness in http adaptive streaming over
	lte network,'' \emph{IEEE Transactions on Circuits and Systems for Video
		Technology}, vol.~26, no.~12, pp. 2284--2298, Dec. 2016.
	
	\bibitem{ITU2004-P880}
	{Recommendation ITU-T P.880}, ``{Methods for objective and subjective
		assessment of quality: Continous evaluation of time varying speech
		quality.}'' \emph{International Telecommunication Union}, 2004.
	
	\bibitem{ITU2012-BT500}
	{Recommendation ITU-R BT.500-13}, ``{Methodology for the subjective assessment
		of the quality of television pictures},'' \emph{International
		Telecommunication Union}, 2012.
	
	\bibitem{QoE_ZWangTimevarying}
	A.~Rehman and Z.~Wang, ``{Perceptual experience of time-varying video
		quality},'' in \emph{2013 Fifth International Workshop on Quality of
		Multimedia Experience (QoMEX)}, Klagenfurt am Worthersee, Austria, 2013, pp.
	218--223.
	
	\bibitem{QoE_Chen2014}
	C.~Chen, L.~K. Choi, G.~{De Veciana}, C.~Caramanis, R.~W. Heath, and A.~C.
	Bovik, ``{Modeling the time-Varying subjective quality of HTTP video streams
		with rate adaptations},'' \emph{IEEE Transactions Image Processing}, vol.~23,
	no.~5, pp. 2206--2221, 2014.
	
	\bibitem{QMon_shafiq_infocom18}
	M.~H. Mazhar and M.~Z. Shafiq, ``{Real-time Video Quality of Experience
		Monitoring for HTTPS and QUIC},'' in \emph{Proceedings of the IEEE INFOCOM},
	Honolulu, HI, USA, 2018.
	
	\bibitem{QMon_seufert2019_stream}
	M.~Seufert, P.~Casas, N.~Wehner, G.~Li, and L.~Kuang, ``Stream-based machine
	learning for real-time qoe analysis of encrypted video streaming traffic,''
	in \emph{3rd International Workshop on Quality of Experience Management
		(QoE-Management)}, Paris, France, 2019.
	
	\bibitem{QMon_wassermann2019_let}
	S.~Wassermann, M.~Seufert, P.~Casas, G.~Li, and L.~Kuang, ``Let me decrypt your
	beauty: Real-time prediction of video resolution and bitrate for encrypted
	video streaming,'' Paris, France, 2019.
	
	\bibitem{texas2018}
	D.~{Ghadiyaram}, J.~{Pan}, and A.~C. {Bovik}, ``Learning a continuous-time
	streaming video qoe model,'' \emph{IEEE Transactions on Image Processing},
	vol.~27, no.~5, pp. 2257--2271, May 2018.
	
	\bibitem{QoE_BampisRecurrent2018ML}
	C.~G. Bampis, Z.~Li, I.~Katsavounidis, and A.~C. Bovik, ``{Recurrent and
		Dynamic Models for Predicting Streaming Video Quality of Experience},''
	\emph{IEEE Transactions on Image Processing}, vol.~27, no.~7, pp. 3316--3331,
	Jul. 2018.
	
	\bibitem{QoE_EswaraLSTM}
	N.~{Eswara}, S.~{Ashique}, A.~{Panchbhai}, S.~{Chakraborty}, H.~P. {Sethuram},
	K.~{Kuchi}, A.~{Kumar}, and S.~S. {Channappayya}, ``Streaming video qoe
	modeling and prediction: A long short-term memory approach,'' \emph{IEEE
		Transactions on Circuits and Systems for Video Technology}, pp. 1--1, 2019.
	
	\bibitem{QoE_bovikTimeVarying2011}
	K.~Seshadrinathan and A.~C. Bovik, ``{Temporal hysteresis model of time varying
		subjective video quality},'' in \emph{2011 IEEE International Conference on
		Acoustics, Speech and Signal Processing (ICASSP)}, Prague, Czech Republic,
	May 2011, pp. 1153--1156.
	
	\bibitem{QoE_bampis2017Dataset}
	C.~G. Bampis, Z.~Li, A.~K. Moorthy, I.~Katsavounidis, A.~Aaron, and A.~C.
	Bovik, ``{Study of Temporal Effects on Subjective Video Quality of
		Experience},'' \emph{IEEE Transactions on Image Processing}, vol.~26, no.~11,
	pp. 5217--5231, Nov. 2017.
	
	\bibitem{qualinet2013qoe}
	P.~Le~Callet, S.~M{\"o}ller, and A.~{Perkis (eds)}, ``{Q}ualinet {W}hite
	{P}aper on {D}efinitions of {Q}uality of {E}xperience,'' Lausanne,
	Switzerland, Tech. Rep., 2013, version 1.2.
	
	\bibitem{QoE_texas2019}
	D.~{Ghadiyaram}, J.~{Pan}, and A.~C. {Bovik}, ``A subjective and objective
	study of stalling events in mobile streaming videos,'' \emph{IEEE
		Transactions on Circuits and Systems for Video Technology}, vol.~29, no.~1,
	pp. 183--197, Jan. 2019.
	
	\bibitem{QoE_tobias2013_DTMA}
	T.~Ho{\ss}feld, R.~Schatz, E.~Biersack, and L.~Plissonneau, ``{Internet Video
		Delivery in YouTube: From Traffic Measurements to Quality of Experience},''
	\emph{Data Traffic Monitoring and Analysis}, vol. 7754, pp. 264--301, 2013.
	
	\bibitem{QoE_singh2012qualityML}
	K.~D. Singh, Y.~Hadjadj-Aoul, and G.~Rubino, ``{Quality of experience
		estimation for adaptive HTTP/TCP video streaming using H. 264/AVC},'' in
	\emph{2012 IEEE Consumer Communications and Networking Conference (CCNC)},
	Las Vegas, USA, Jan. 2012, pp. 127--131.
	
	\bibitem{QoE_liu2015_deriving}
	Y.~Liu, S.~Dey, F.~Ulupinar, M.~Luby, and Y.~Mao, ``{Deriving and validating
		user experience model for DASH video streaming},'' \emph{IEEE Transactions on
		Broadcasting}, vol.~61, no.~4, pp. 651--665, 2015.
	
	\bibitem{QoE_rodriguez2016_video}
	D.~Z. Rodr{\'\i}guez, R.~L. Rosa, E.~C. Alfaia, J.~I. Abrah{\~a}o, and
	G.~Bressan, ``{Video quality metric for streaming service using DASH
		standard},'' \emph{IEEE Transactions on Broadcasting}, vol.~62, no.~3, pp.
	628--639, 2016.
	
	\bibitem{tran2016_GC}
	H.~T.~T. Tran, N.~P. Ngoc, A.~T. Pham, and T.~C. Thang, ``{A Multi-Factor QoE
		Model for Adaptive Streaming over Mobile Networks},'' in \emph{2016 IEEE
		Globecom Workshops (GC Wkshps)}, Washington DC, USA, Dec. 2016, pp. 1--6.
	
	\bibitem{QoE_tavakoli2016_JSAC}
	S.~Tavakoli, S.~Egger, M.~Seufert, R.~Schatz, K.~Brunnstr{\"o}m, and
	N.~Garc{\'\i}a, ``{Perceptual quality of HTTP adaptive streaming strategies:
		Cross-experimental analysis of multi-laboratory and crowdsourced subjective
		studies},'' \emph{IEEE Journal on Selected Areas in Communications}, vol.~34,
	no.~8, pp. 2141--2153, 2016.
	
	\bibitem{QoE_bellLab2013_QoEmodel}
	J.~D. Vriendt, D.~D. Vleeschauwer, and D.~Robinson, ``{Model for estimating QoE
		of video delivered using HTTP adaptive streaming},'' in \emph{2013 IFIP/IEEE
		International Symposium on Integrated Network Management (IM 2013)}, Ghent,
	Belgium, May 2013, pp. 1288--1293.
	
	\bibitem{QoE_ywang2015_assessing}
	Z.~Guo, Y.~Wang, and X.~Zhu, ``Assessing the visual effect of non-periodic
	temporal variation of quantization stepsize in compressed video,'' in
	\emph{2015 IEEE International Conference on Image Processing (ICIP)}, Quebec
	City, Canada, Sept. 2015, pp. 3121--3125.
	
	\bibitem{QoE_tobias2014_assessing}
	T.~Ho{\ss}feld, M.~Seufert, C.~Sieber, and T.~Zinner, ``{Assessing effect sizes
		of influence factors towards a QoE model for HTTP adaptive streaming},'' in
	\emph{2014 Sixth International Workshop on Quality of Multimedia Experience
		(QoMEX)}, Singapore, Sept. 2014, pp. 111--116.
	
	\bibitem{QoE_Seufert2013_pool}
	M.~Seufert, M.~Slanina, S.~Egger, and M.~Kottkamp, ``{``To pool or not to
		pool'': A comparison of temporal pooling methods for HTTP adaptive video
		streaming},'' in \emph{5th International Conference Quality Multimedia
		Experience}, Klagenfurt am Worthersee, Austria, Jul. 2013, pp. 52--57.
	
	\bibitem{ITU1203_3}
	{Recommendation ITU-T P.1203.3}, ``{Parametric bitstream-based quality
		assessment of progressive download and adaptive audiovisual streaming
		services over reliable transport-Quality integration module},''
	\emph{International Telecommunication Union}, 2017.
	
	\bibitem{tran2018_QoMEX}
	H.~T.~T. Tran, N.~P. Ngoc, T.~Ho{\ss}feld, and T.~C. Thang, ``{A Cumulative
		Quality Model for HTTP Adaptive Streaming},'' in \emph{2018 Tenth
		International Conference on Quality of Multimedia Experience (QoMEX)},
	Sardinia, Italy, May 2018, pp. 1--6.
	
	\bibitem{thang2013_JCN}
	T.~C. Thang, H.~T. Le, H.~X. Nguyen, A.~T. Pham, J.~W. Kang, and Y.~M. Ro,
	``{Adaptive video streaming over HTTP with dynamic resource estimation},''
	\emph{Journal of Communications and Networks}, vol.~15, no.~6, pp. 635--644,
	2013.
	
	\bibitem{HAS_muller2012_evaluation}
	C.~M{\"u}ller, S.~Lederer, and C.~Timmerer, ``{An evaluation of dynamic
		adaptive streaming over HTTP in vehicular environments},'' in
	\emph{Proceedings of the 4th Workshop on Mobile Video}, Chapel Hill, North
	Carolina, Feb. 2012, pp. 37--42.
	
	\bibitem{QoE_nam2016_qoewhycat}
	H.~Nam, K.-H. Kim, and H.~Schulzrinne, ``{QoE matters more than QoS: Why people
		stop watching cat videos},'' in \emph{35th Annual IEEE International
		Conference on Computer Communications}, San Francisco, CA, USA, Apr. 2016,
	pp. 1--9.
	
	\bibitem{P.9132014}
	{Recommendation ITU-T P.913}, ``{Methods for the subjective assessment of video
		quality, audio quality and audiovisual quality of Internet video and
		distribution quality television in any environment},'' \emph{International
		Telecommunication Union}, 2014.
	
	\bibitem{QoE_Kahneman1993}
	D.~Kahneman, B.~L. Fredrickson, C.~A. Schreiber, and D.~A. Redelmeier, ``{When
		More Pain Is Preferred to Less: Adding a Better End},'' \emph{Psychological
		Sciences}, vol.~4, no.~6, pp. 401--405, Nov. 1993.
	
	\bibitem{QoE_Koster2017}
	F.~K{\"o}ster, G.~Mittag, and S.~M{\"o}ller, ``{Modeling the overall quality of
		experience on the basis of underlying quality dimensions},'' in \emph{9th
		International Conference Quality Multimedia Experience}, Erfurt, Germany, May
	2017, pp. 1--6.
	
	\bibitem{QoE_Yin_2015}
	X.~Yin, A.~Jindal, V.~Sekar, and B.~Sinopoli, ``{A control-theoretic approach
		for dynamic adaptive video streaming over HTTP},'' \emph{ACM SIGCOMM Computer
		Communications Review}, vol.~45, no.~4, pp. 325--338, 2015.
	
	\bibitem{ITUT_implement1}
	A.~Raake, M.-N. Garcia, W.~Robitza, P.~List, S.~G{\"{o}}ring, and B.~Feiten,
	``{A bitstream-based, scalable video-quality model for HTTP adaptive
		streaming: ITU-T P.1203.1},'' in \emph{Ninth International Conference on
		Quality of Multimedia Experience (QoMEX)}, Erfurt, Germany, May 2017, pp.
	1--6.
	
	\bibitem{ITUT_implement2}
	W.~Robitza, S.~G{\"{o}}ring, A.~Raake, D.~Lindegren, G.~Heikkil{\"{a}},
	J.~Gustafsson, P.~List, B.~Feiten, U.~W{\"{u}}stenhagen, M.-N. Garcia,
	K.~Yamagishi, and S.~Broom, ``{HTTP Adaptive Streaming QoE Estimation with
		ITU-T Rec. P.1203 - Open Databases and Software},'' in \emph{Proceedings of
		the 9th ACM Multimedia Systems Conference}, Amsterdam, Netherlands, Jun.
	2018, pp. 466--471.
	
	\bibitem{ITUT_implement3}
	{Recommendation ITU-T P.1203}, ``{ITU-T Rec. P.1203 Standalone
		Implementation},'' 2018, \url{https://github.com/itu-p1203/itu-p1203/},
	accessed 2018-07-01.
	
	\bibitem{QoE_Database_ZDuanmu2018}
	Z.~Duanmu, A.~Rehman, and Z.~Wang, ``{A Quality-of-Experience Database for
		Adaptive Video Streaming},'' \emph{IEEE Transactions on Broadcasting},
	vol.~64, no.~2, pp. 474--487, Jun. 2018.
	
	\bibitem{QoE_QoEIndex_ZDuanmu2018}
	Z.~Duanmu, K.~Zeng, K.~Ma, A.~Rehman, and Z.~Wang, ``{A Quality-of-Experience
		Index for Streaming Video},'' \emph{IEEE Journal of Selected Topics in Signal
		Processing}, vol.~11, no.~1, pp. 154--166, Feb 2017.
	
	\bibitem{ITUT_Rec1401}
	{Recommendation ITU-T P.1401}, ``{Methods, metrics and procedures for
		statistical evaluation, qualification and comparison of objective quality
		prediction models },'' \emph{International Telecommunication Union}, 2012.
	
	\bibitem{tran2017_IEICEhistogram}
	H.~T.~T. Tran, N.~P. Ngoc, Y.~J. Jung, A.~T. Pham, and T.~C. Thang, ``{A
		Histogram-Based Quality Model for HTTP Adaptive Streaming},'' \emph{IEICE
		Transactions on Fundamentals of Electronics, Communications and Computer
		Sciences}, vol. 100, no.~2, pp. 555--564, 2017.
	
\end{thebibliography}
% Generated by IEEEtran.bst, version: 1.14 (2015/08/26)

\end{document}